\documentclass[aps,prb,amsfonts,amssymb,twocolumn,amsmath,preprintnumbers,nofootinbib,superscriptaddress,floatfix,showpacs]{revtex4}
\usepackage[dvips]{color}
\usepackage[dvips]{graphics}
\usepackage{graphicx}

\def\L{{\cal L}}
\def\vpf{{\vec p_f}}
\def\vvf{{\vec v_f}}
\def\cN{{\cal N}}
\def\cNf{{{\cal N}_f}}

\def\vp{{\mbox{\boldmath$p$}}}

\newcommand{\gRo}{\hat g^R_0}
\newcommand{\gRd}{g^R_3}

\newcommand{\gK}{\hat g^K}

\newcommand{\grad}{\mbox{\boldmath$\nabla$}}

\newcommand{\sech}{\mbox{sech}}

\begin{document}

\title{Large Thermoelectric Effects in Unconventional Superconductors}

\author{T. L\"ofwander}
\email{tomas@snowmass.phys.nwu.edu}
\affiliation{Department of Physics \& Astronomy,
Northwestern University, Evanston, IL 60208}

\author{M.~Fogelstr\"om}
\email{mikael.fogelstrom@mc2.chalmers.se}
\affiliation{Applied Quantum Physics, MC2, Chalmers,
S-41296 G\"oteborg, Sweden}

\date{\today}

\begin{abstract}
We present analytic and numerical results for the thermoelectric
effect in unconventional superconductors with a dilute random
distribution of impurities, each scattering isotropically but with a
phase shift intermediate between the Born and unitary limits. The
thermoelectric response function has a linear temperature dependence
at low temperatures, with a slope that depends on the impurity
concentration and phase shift.  Although the thermoelectric effect
vanishes identically in the strict Born and unitary limits, even a
small deviation of the phase shift from these limits leads to a large
response, especially in clean systems. We also discuss possibilities
of measuring counter-flowing supercurrents in a SQUID-setup. The
non-quantized thermoelectrically induced flux can easily be of the
order of a percent of the flux quantum in clean systems at $^4{\rm
He}$ temperatures.
\end{abstract}


\maketitle

\section{Introduction}

Electronic charge and heat transport measurements can give important
information about the microscopic properties of high-$T_c$ and other
unconventional
superconductors.\cite{bonn92,tai97,huss99,chi00,cor00,nak01,hill01,and02,suz02,pro02,sut03,hill04,tur03,pet86,hir88,lee93,hir93,graf96,hir96,kim03}
Of particular interest is the low-temperature regime, where electronic
properties are believed to be controlled by elastic impurity
scattering. This is the case, because in an unconventional
superconductor with an order parameter changing sign at some points on
the Fermi surface (for example the $d$-wave order parameter), elastic
scattering by even a small concentration of impurities leads to pair
breaking, and the formation of an impurity band of width $\gamma$
around the Fermi level. These low-lying excitations, are responsible
for a {\it universal} limit at low-temperatures $T\ll\gamma$, where
the charge and heat conductances become independent of the scattering
properties of the impurities. This low-energy behavior was first
predicted by Lee\cite{lee93} for the zero-frequency charge
conductance, and was later extended to include the heat conductance by
Graf {\it et al.}.\cite{graf96}

The universal character of the $T\rightarrow 0$ heat conductance was
seen experimentally in
Refs.~[\onlinecite{tai97,chi00,nak01,and02,pro02,sut03}]. A series of
recent related experiments on the temperature dependence of the heat
conductance approaching the universal limit,\cite{hill04} microwave
conductivity,\cite{tur03} and STM spectroscopy of quasiparticle
impurity states,\cite{pan00} have all indicated that impurity
scattering in the cuprates might not be in the strict unitary or Born
limits. Limiting the discussion to isotropic scattering, this means
that the impurity potential $u_0$ is of intermediate size, and the
corresponding $s$-wave scattering phase shift
$\delta_0=\arctan(\pi\cN_fu_0)$ has some intermediate value
$0<\delta_0<\pi/2$. Here is $\cN_f$ the density of states at the Fermi
level in the normal state. If this is the case, electron-hole symmetry
is explicitly broken near each impurity and the global electron-hole
symmetry is broken for a homogeneous dilute distribution of such
impurities.\cite{monien87b,arf88,arf89,sal96} This has direct
consequences for the heat and charge transport
coefficients,\cite{graf96} but leads also to large thermoelectric
effects. A fact that was previously noted in connection to the heavy
fermion systems by Arfi {\it et al.}\cite{arf88,arf89} In the present
paper we report an extensive analysis of electron-hole symmetry
breaking by elastic impurity scattering, and its effect on the
transport coefficients in unconventional superconductors, with an
emphasis on the d-wave cuprates at low temperatures.

Previous work on thermoelectric effects in the heavy fermion
superconductors\cite{arf88,arf89,hir88b} were limited to situations in
which the energy-dependent broadening of the quasiparticle states can
be neglected, i.e. when the imaginary part of the impurity self-energy
satisfies $\Im \Sigma^R_3(\epsilon)\ll\epsilon$. This approximation is
expected to be good at high temperatures, $T<T_c$, but fails at lower
temperatures. In fact, in the temperature regime $T\alt\gamma$ of high
current interest, where effects of universality become of importance,
the impurity renormalization is the dominant energy scale and the
problem has to be considered anew.

Our results can be summarized as follows. (1) For intermediate phase
shifts, the thermoelectric response function $\eta(T)$ is in general
large, which leads to counter-flowing supercurrents detectable e.g. as
a thermally induced magnetic flux in a ring
setup.\cite{gin44,gal74,har74,zav74,har80} (2) At low temperatures
$T\alt\gamma$, $\eta(T)$ scales linearly with temperature, with a
non-universal slope that grows large in {\it clean} systems.

\section{Electron-hole symmetry breaking and giant thermoelectric effects}

To compute the response to a thermal gradient $\nabla T$ and an
electric field ${\vec E}(\omega)$, we need to evaluate the charge
current $\delta \vec j_e$ as well as the heat current $\delta \vec
j_\epsilon$. In the linear response, the observable response functions
such as the heat and charge conductivities, can conveniently be
expressed in terms of response functions $\L_{\alpha\beta}$ that are
defined as
\begin{equation}
\left(\begin{array}{c}\delta \vec j_e \\ \delta \vec j_\epsilon \end{array}\right)=
\left(\begin{array}{cc} \tensor\L_{11}& \tensor\L_{12}\\ \tensor\L_{21} & \tensor\L_{22} \end{array}\right) 
\left(\begin{array}{c} \frac{\vec E}{T} \\ \nabla (\frac{1}{T}) \end{array}\right).
\label{Onsager}
\end{equation}
This particular choice of forces ensures that the Onsager relations
$\L_{12}=\L_{21}$ are fulfilled (see e.g. Ref.~[\onlinecite{mahan}]).

Contrary to the two diagonal terms $\L_{11}$ and $\L_{22}$ in
Eq.~(\ref{Onsager}), the thermoelectric coefficients $\L_{12}=\L_{21}$
require an electron-hole asymmetry around the Fermi energy in order to
be non-zero. In the normal state, to quasiclassical accuracy, i.e. to
leading order in small parameters ${\sf s}=\{T/T_f,1/p_f\xi_0,...\}$,
there is no such electron-hole asymmetry in the theory and the
thermoelectric coefficients vanish, $\L_{12}^N=\L_{21}^N=0$. Here is
$\xi_0=v_f/T_c$ the superconducting coherence length, and $T_f$,
$p_f$, and $v_f$ are the Fermi temperature, momentum and velocity,
respectively. This result holds also in the superconducting phase when
the order parameter has the conventional $s$-wave
symmetry.\cite{gal74} However, for unconventional superconductors this
is not the case: impurity scattering is pair-breaking and potential
scattering off an impurity site induces a non-vanishing electron-hole
asymmetry already to leading order in the small parameters $\sf{s}$
[\onlinecite{monien87b,arf88,arf89,sal96}]. To illustrate this, we
consider a superconductor having an order parameter with a vanishing
Fermi surface average, $\int d\vpf \Delta(\vpf)=0$, in which case the
equilibrium-state impurity $\hat t^R_{\rm imp}$-matrix in Nambu
(electron-hole) space has the form
\begin{equation}
\hat t^R_{\rm imp}(\epsilon) = \frac{\sin\delta_0}{\pi \cNf}
\frac{\cos\delta_0 \hat 1 + \sin \delta_0\int d\vpf \gRd(\vpf,\epsilon)\hat\tau_3}
{\cos^2\!\delta_0-\sin^2\!\delta_0[\int d\vpf \gRd(\vpf,\epsilon)]^2},
\label{t-matrix}
\end{equation}
where the diagonal component of the equilibrium quasiclassical Green's
function is $\gRd=-\tilde \epsilon^R /\Omega^R$
[$\Omega^R=\sqrt{|\Delta(\vpf)|^2 -(\tilde \epsilon^R)^2}$]. The
electron-hole asymmetric scattering becomes explicit when we examine
the electron (11) and hole (22) parts:
\begin{equation}
t^R_{11 (22)}=\frac{1}{
\cos\delta_0\mp\sin\delta_0\int d\vpf
\gRd(\vpf,\epsilon)}.
\label{t-matrix-poles}
\end{equation}
In particular, multiple scattering off the impurity leads to
resonances at $\epsilon=-\epsilon_{\rm res}(\delta_0)$ for
electron-like excitations and at $\epsilon=+\epsilon_{\rm
res}(\delta_0)$ for hole-like excitations. These impurity resonances
become virtual bound states localized at the impurity in the strong
scattering limit, $\delta_0\rightarrow\pi/2$ [\onlinecite{bal95}].
This also implicates that electrons and holes have different energy
dependent scattering life times at intermediate phase
shifts.\cite{monien87b}

\begin{figure}[t]
\includegraphics[width=8cm]{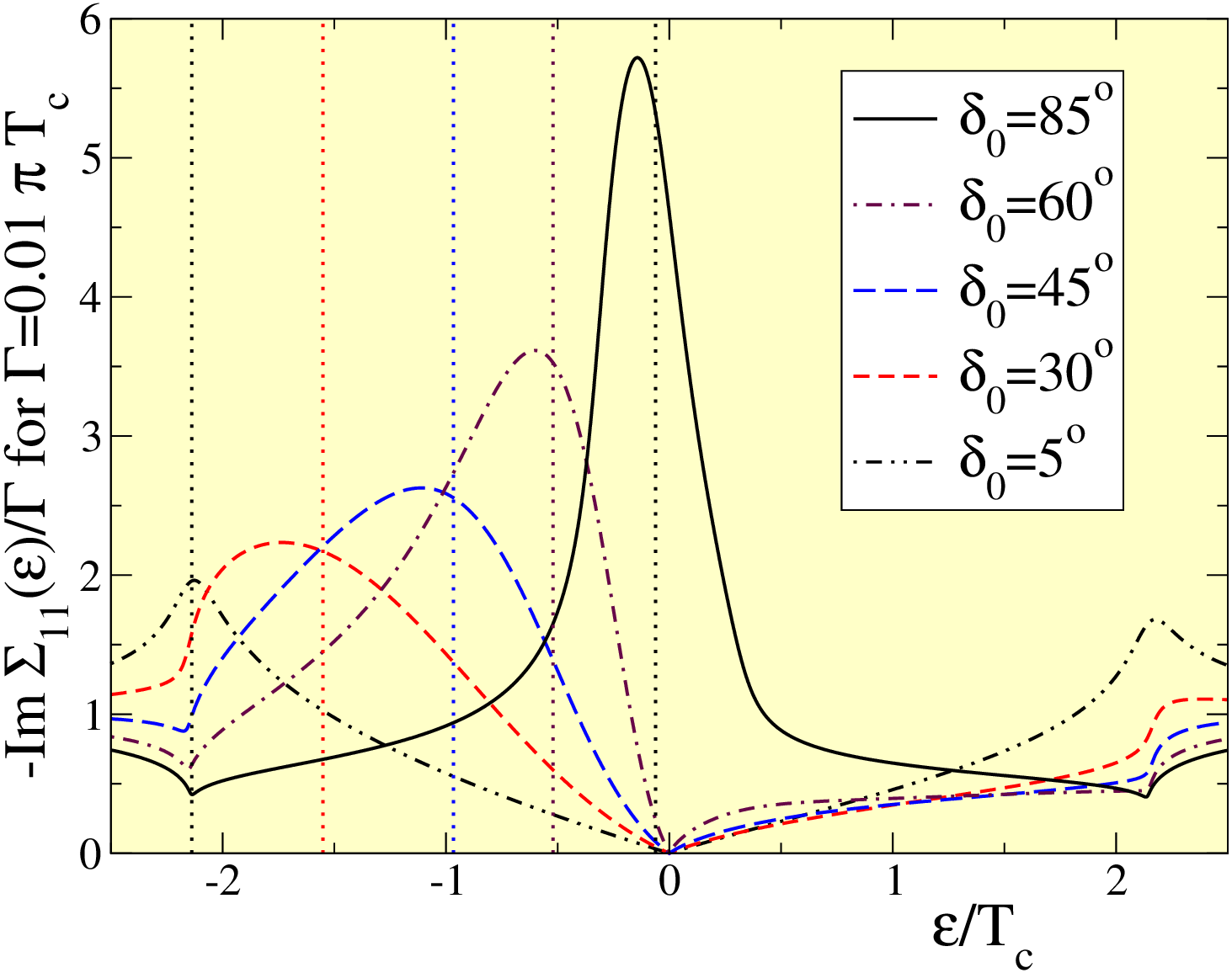}
\caption{The imaginary part of $\Sigma^R_{11}$ shown for different
normal state scattering phase shifts $\delta_0$ for a $d$-wave
superconductor with $\Delta(\vpf)=\Delta_0 \cos(2\phi_{\vpf})$. There
is a substantial asymmetry in $\Im \Sigma^R_{11}(\epsilon)$ around
$\epsilon=0$. The dashed vertical lines indicate the single-impurity
resonance energies $\epsilon_{\rm res}(\delta_0)$ for an electron-like
quasiparticle at the corresponding phase shifts. The resonance energy
of the single impurity self energy is carried over to the
self-consistently computed self energy describing the randomly
distributed impurities, but slightly shifted away from the Fermi
level: $\bar \epsilon(\delta_0)\alt \epsilon_{\rm res}(\delta_0)$. The
hole component, $\Sigma^R_{22}$, is related to the electron component
as
$\Sigma^R_{22}(\epsilon)=\Sigma^R_{0}-\Sigma^R_{3}=[\Sigma^R_{11}(-\epsilon)]^*$.
}\label{fig1_Resonances}
\end{figure}

We assume that the impurities are randomly distributed with an average
small concentration $n_i$, that is absorbed in the normal state
elastic scattering rate $\tau^{-1}_0=2\Gamma=2 n_i \sin^2\delta_0/\pi
\cNf$. Within the usual impurity averaging technique\cite{AGD} the
impurity self energy has the form
\begin{equation}
\hat \Sigma^R_{\rm imp}(\epsilon)=
\Sigma^R_{0}(\epsilon) \hat 1 + \Sigma^R_{3}(\epsilon) \hat\tau_3 =
n_i \hat t^R_{\rm imp}(\epsilon).
\label{eq:sR0}
\end{equation}
The term proportional to the third
Pauli matrix in electron-hole space, $\hat\tau_3$, gives the impurity
renormalization of the energy, $\tilde
\epsilon^R=\epsilon-\Sigma_3^R$. The unit term,
$\Sigma^R_{0}(\epsilon)$, drops out of the equations for the
equilibrium state, but in fact enters the transport equations
describing the non-equilibrium state. $\Sigma^R_{0}(\epsilon)$
explicitly breaks electron-hole symmetry, which leads to the
thermoelectric effect we study below. In general we can not expect the
unit term to either vanish or to be small, although it vanishes in the
strict Born and unitary limits. The induced asymmetry is shown for the
electron part $\Sigma^R_{11}=\Sigma^R_{0}+\Sigma^R_{3}$ in
Fig.~\ref{fig1_Resonances}.

The response functions $\L_{\alpha\beta}$, calculated to lowest order
in the small parameters ${\sf s}$, are conveniently computed through
the quasiclassical Keldysh propagator $\delta\gK$, evaluated to linear
order in the forces. Amazingly, $\delta\gK$ has a closed form in which
the self-consistently computed equilibrium Green's function $\gRo$,
order parameter $\Delta^R$, and impurity self-energies
$\hat\Sigma^R_{\rm imp}$ serve as input.\cite{RS,graf96} A
complication is that $\delta\gK$ depends on linear corrections to the
self-energies $\delta\hat\sigma^R$ and $\delta\hat\sigma^K$, which
need to be computed self-consistently with $\delta\gK$. This procedure
is equivalent to take into account vertex corrections in the Kubo
formalism. In our model, we assume isotropic scattering, in which case
the vertex corrections vanish. This is a well known fact, and is
ultimately due to the anti-symmetric form of the forces
$\vvf\cdot{\vec E}$ and $\vvf\cdot\nabla T$. For cases of anisotropic
impurity scattering ($p$-wave etc.), these corrections must be taken
into account, as was done for the diagonal response functions
$\L_{11}$ and $\L_{22}$ by Durst and Lee.\cite{durst00} In the
following we neglect anisotropic scattering and compute all response
functions $\L_{\alpha\beta}$.

The expression for $\delta\gK$ given in the appendix of
Ref.~[\onlinecite{graf96}] is general enough to serve as the starting
point for our calculation. In fact, the calculation of the
zero-frequency limit of the off-diagonal response functions
$\L_{12}=\L_{21}$ follows the same logical steps as the calculation of
$\L_{11}$ and $\L_{22}$ in Ref.~[\onlinecite{graf96}] [their Eqs.~(29)
and (30)], and we only give the final result:
\begin{widetext}
\begin{equation}
\L_{12}^{ij} = - \frac{e}{4}
\int d\epsilon\, \epsilon\, {\sech}^2\frac{\epsilon}{2 T} 
\int d\vpf [v_{f,i} v_{f,j}]
\frac{\cN(\vpf,\epsilon)\,\Im\,\Sigma^R_{0}(\epsilon)}
{\left[\Re\,\Omega^R(\vpf;\epsilon)\right]^2-\left[\Im\,\Sigma^R_{0}(\epsilon)\right]^2},
\label{eq:L12}\end{equation}
\end{widetext}
where ${\cal N}(\vpf,\epsilon)= - \cN_f \Im\,\left[\tilde
\epsilon^R/\Omega^R(\vpf;\epsilon)\right]$ is the density of states in
the superconducting phase. $\L_{12}$ is directly proportional to the
imaginary part of the unit term $\Im\,\Sigma^R_0$ of the equilibrium
impurity self energy, which is an odd function of
energy. Eq.~(\ref{eq:L12}) is the proper generalization to arbitrary
low temperatures, including impurity scattering renormalization
effects, of the results in
Refs.~[\onlinecite{arf88}-\onlinecite{arf89}].

\subsection{Conductivities}

Once we know $\L^{ij}_{\alpha\beta}$, we may compute the
conductivities. In the following we consider transport in the cuprate
planes, along one of the anti-nodes of the $d$-wave order
parameter. We can then drop the superscripts of
$\L^{ij}_{\alpha\beta}$. The charge and heat conductances are given by
\begin{equation}\begin{split}
\sigma(T) = \frac{\L_{11}(T)}{T},\\
\kappa(T) = \frac{\L_{22}(T)}{T^2}.
\end{split}\end{equation}
In the presence of a temperature gradient, the non-vanishing
thermoelectric coefficient $\L_{12}$ induces a charge current $\delta
j_e=-\eta\grad T$, where we define
\begin{equation}
\eta(T)=\frac{\L_{12}(T)}{T^2}.
\end{equation}
The appearence of a bulk charge current, $\delta j_e$,
leads to a counter-flowing supercurrent, which we discuss in the
next section.
\begin{figure}[b]
\includegraphics[width=8cm]{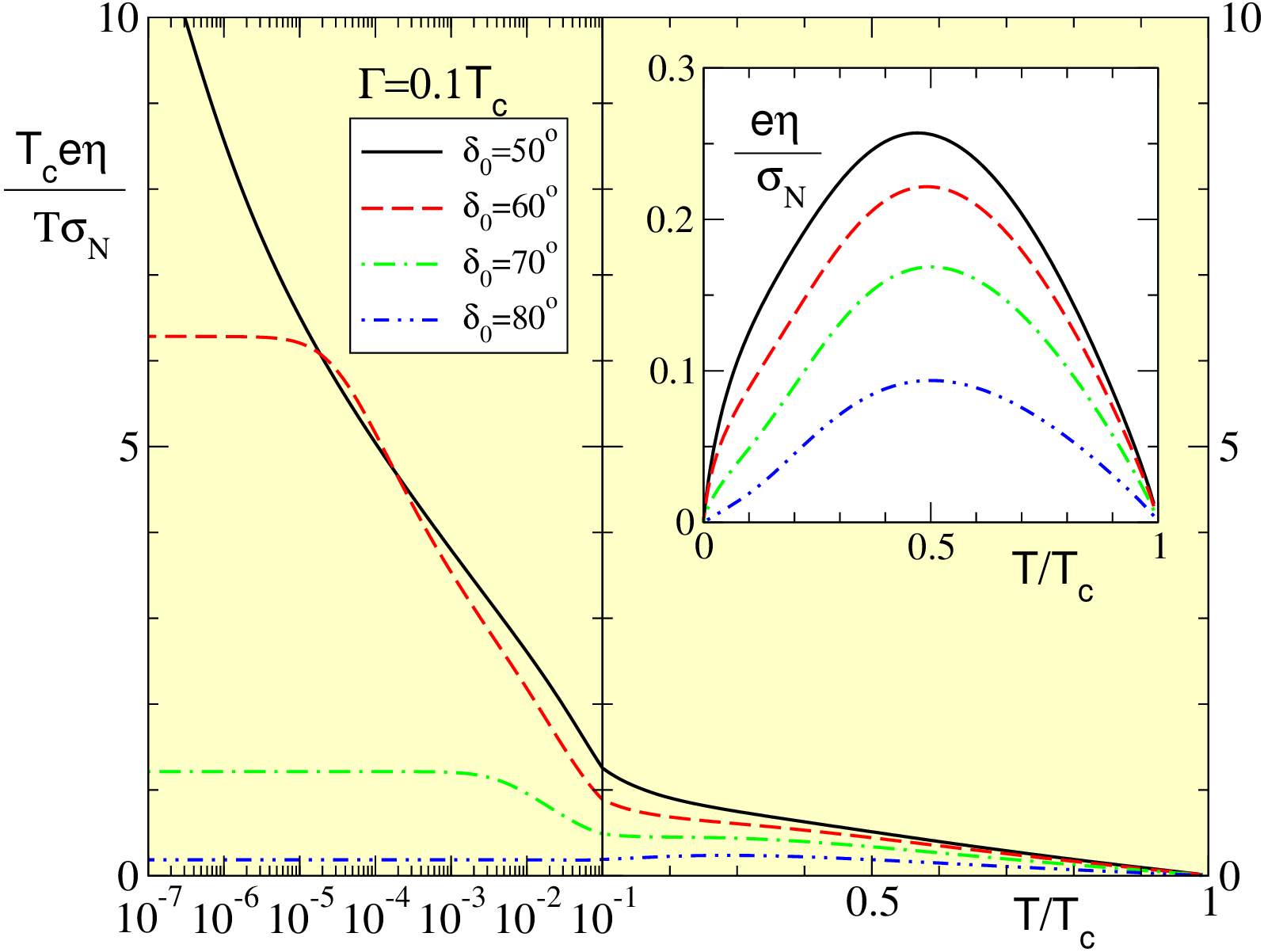}
\caption{The thermoelectric coefficient $\eta$ scaled by its low-$T$
dependence $\eta\sim T$ for the scattering rate $\Gamma=0.1T_c$
and several normal state scattering cross sections
$\sigma_i=\sin^2\delta_0$. The inset shows the unscaled function. In
the {\it universal} limit ($T\ll\gamma$), $\eta/T$ approaches a {\it
non-universal} constant given in Eq.~(\ref{eq:eta-lowT}).}
\label{fig2:eta}
\end{figure}

In Fig.~\ref{fig2:eta} we present the thermoelectric response function
$\eta$ as a function of temperature for several phase shifts for a
fixed rather short mean free path $\ell=v_f\tau_0\sim 5\xi_0$, where
$\xi_0=v_f/T_c$ is the superconducting coherence length. In our model,
the physical transition temperature $T_c$ is then suppressed by about
$25\%$ compared to the clean limit transition temperature $T_{c0}$, in
accordance with the Abrikosov-Gorkov form for the $T_c$-suppression as
function of $\Gamma=(1/2\tau_0)$ by elastic impurity
scattering.\cite{xu95} $\eta(T)$ is sizable over a wide temperature
range from zero to $T_c$, but vanishes in the $T\rightarrow T_c$
limit, where the electron-hole asymmetry is of order ${\sf s}^2$ and
neglected in our quasiclassical theory. The thermoelectric effect is
sensitive to the microscopic superconducting properties, such as the
order parameter size and its symmetry, through the coherence factors
entering Eq.~(\ref{eq:L12}). It is also sensitive to the nature of the
impurity scattering: first to the normal state mean free path, but
also to the phase shift $\delta_0$. In the strict Born
($\delta_0\rightarrow0$) and unitary ($\delta_0\rightarrow\pi/2$)
limits, $\eta(T)$ vanishes since the electron-hole asymmetry vanishes
in those limits, $\Im\Sigma^R_0(\epsilon)\equiv 0$. However, for all
other values of $\delta_0$, $\eta(T)$ is large, and has a maximum of
the order of $0.1 \sigma_N/e$ near $T\sim 0.5 T_c$. Here is
$\sigma_N=e^2\cNf v_f^2 /(2\Gamma)$ the Drude conductivity in the
normal state. In fact, even a small deviation of the phase shift from
e.g. the unitary limit leads to dramatic changes in the thermoelectric
response, see below. At low temperature, we find $\eta\propto T$, (see
left half of Fig.~\ref{fig2:eta}). The slope $\eta/T$ contains
detailed information about material parameters such as the scattering
phase shift, as we will discuss in more detail in
Section~\ref{sec:lowT}.

We should mention that we have neglected inelastic scattering in our
calculations. In e.g. Ref.~[\onlinecite{hir96}] inelastic scattering
by anti-ferromagnetic spin fluctuations were included in the
calculation of the thermal conductivity, and it was shown to give rise
to the characteristic increase in $\kappa(T)$ seen in experiments
(e.g. [\onlinecite{sut03,hill04}]) just below $T_c$. Thus, we expect
corrections to our results in Fig.~\ref{fig2:eta} in the high
temperature region $T\alt T_c$. In the important low-$T$ region,
$T\sim\gamma$, inelastic scattering is not of importance, since the
corresponding self energy scales as $T^3$ and is small compared to the
self energy from elastic impurity scattering.

\begin{figure}[t]
\includegraphics[width=8cm]{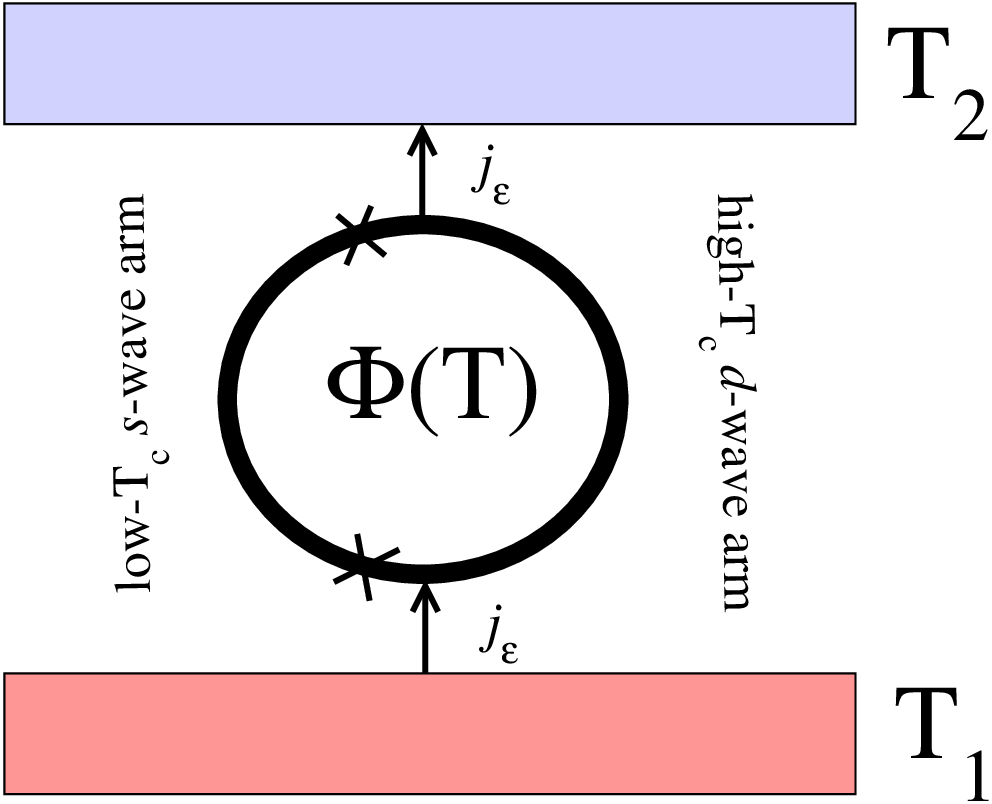}
\caption{A sketch showing the major features needed for the detection
of thermoelectrically induced magnetic fluxes. The loop consists of
two arms, one being an $s$-wave superconductor for which $\eta$ is
orders of magnitude smaller than in the $d$-wave arm. The two
superconductors are connected by Josephson junctions and the $d$-wave
arm is connected to two bulky regions were the temperature is
regulated.}
\label{fig3:exp_setup}
\end{figure}

\subsection{Thermoelectrically generated magnetic flux}

In the normal state, the thermoelectric effect leads to the appearance
of a voltage (given by the thermopower). However, in the
superconducting state, a steady state voltage, and an associated
electric field, is short circuited by the appearance of a supercurrent
and an associated phase gradient,\cite{gin44,gal74,har74}
\begin{equation}
j_s = (e/m) n_s p_s = -\delta j_e,
\end{equation}
where $n_s$ is the superfluid density and
$p_s=\frac{1}{2}(\nabla\chi-\frac{2e}{c}{\bf A})$ is the superfluid
momentum. Thus, the total charge current is zero, $j_e^{\rm
total}=\delta j_e+j_s=0$, but the phase gradient can be detected in a
flux measurement as was done in Ref.~[\onlinecite{zav74,har80}]. These
early experiments were carried out with low-$T_c$ $s$-wave
superconductors, for which thermoelectric effects are smaller than in
the unconventional superconductors we are considering, for two
reasons. First, they are down by the electron-hole asymmetry factor
${\sf s}$; second, they are exponentially suppressed at low
temperatures by the energy gap around the Fermi surface. Consequently,
the experiments were mainly done in the temperature region near $T_c$,
where the temperature dependence of $n_s$ plays an important and
somewhat parasitic role when the goal is to measure $\eta(T)$ (for a
recent discussion of the experimental situation see
Ref.~[\onlinecite{gal02}]).

\begin{figure}[t]
\includegraphics[width=8cm]{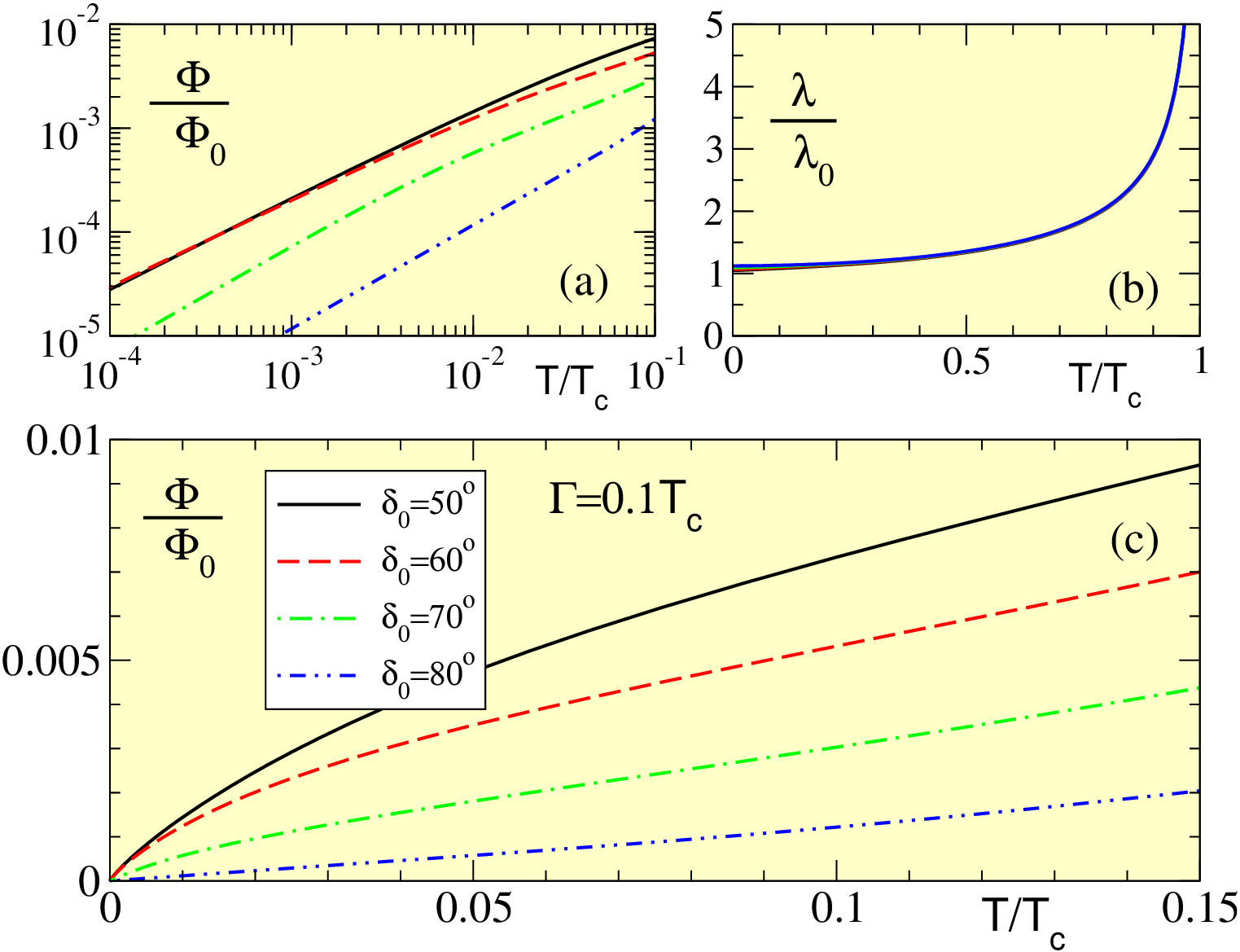}
\caption{The magnetic flux induced by a temperature gradient in a ring
with a thermoelectric response given in Fig.~\ref{fig2:eta}. It is
assumed that a temperature difference $\delta T=T_1-T_2=0.005T_c$ is
maintained over the loop in Fig.~\ref{fig3:exp_setup}. Current
experimental precision for flux measurements is
$10^{-6}\Phi_0/\sqrt{{\rm Hz}}$. Part (b) shows the divergence of the
penetration depth in the $d$-wave superconductor at temperatures near
$T_c$.}
\label{fig4:Phi}
\end{figure}
\begin{figure*}[t]
\includegraphics[width=16cm]{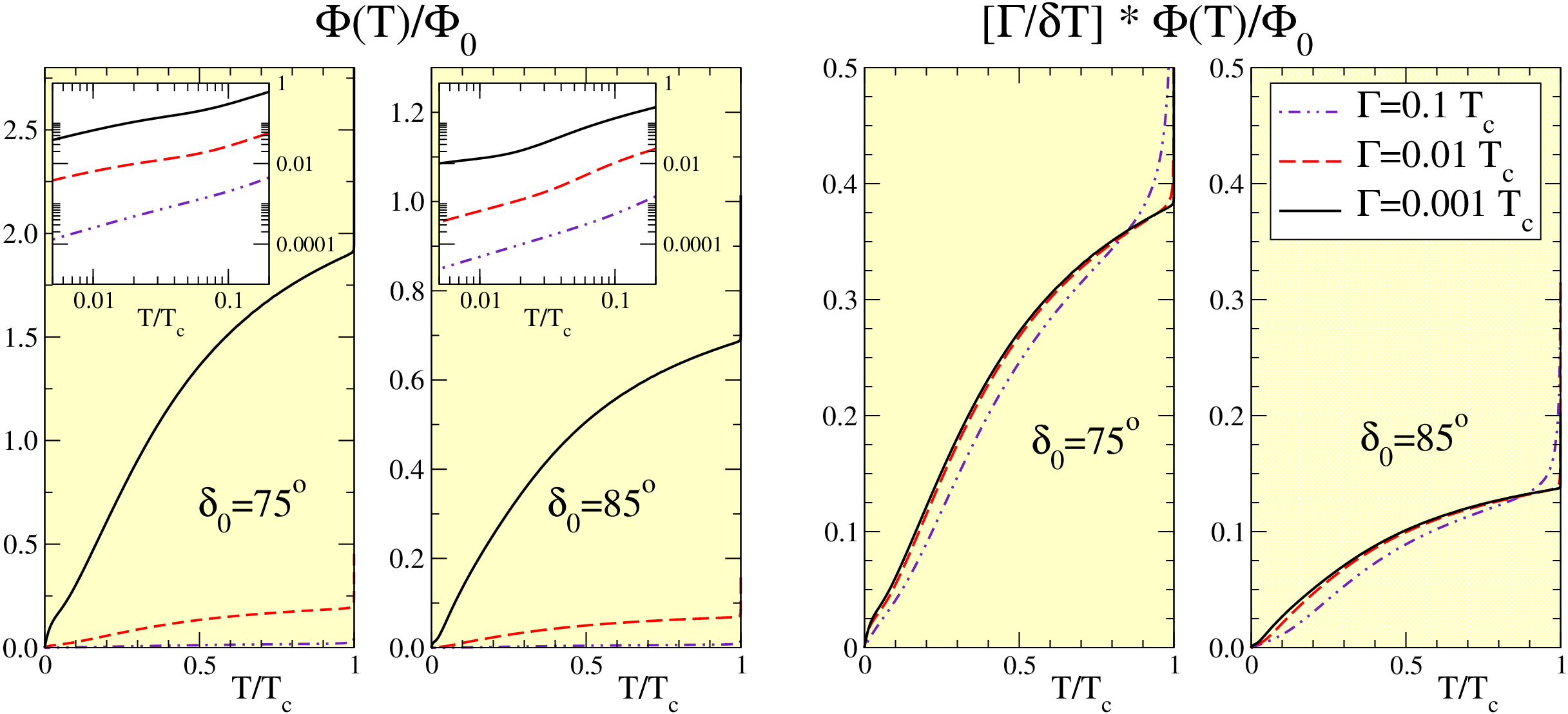}
\caption{The thermoelectrically induced flux as a function of
temperature for strong scattering impurity potentials,
$\delta_0=75^\circ (\sigma_i\approx0.93)$ and $\delta_0=85^\circ
(\sigma_i\approx0.99)$, at different scattering rates (the
corresponding mean-free-paths are: $5, 50$, and $500\,\xi_o$).  In the
two left panels we show the actual flux generated as a function of
temperature, setting $\delta T=0.005T_c$. At high temperatures, $T\alt
T_c$, the flux is a sizable fraction of $\Phi_0$ and a larger $\delta
T$ may well give a flux $\agt\,\Phi_0$. At low temperatures $T\alt 0.1
T_c$ the flux may still be of order $0.1\%-1\%\, \Phi_0$ if the
material is clean, $\ell \agt 20\, \xi_0$, as indicated in the insets.
In the two panels to the right we show the factor
$(\lambda(T)/\lambda_0)^2 e\eta(T)/\sigma_N$.  This factor depends
strongly on the scattering phase shift $\delta_0$ (see also
Fig.~\ref{fig2:eta} where $e\eta(T)/\sigma_N$ is displayed for a
larger variety of scattering phase shifts) but does not have a
particularly strong dependence on $\Gamma$ apart from close to $T_c$
where the temperature dependence of $\lambda(T)$ dominates.  }
\label{fig5:Phi_strong}
\end{figure*}

To measure the flux generated by the counter-flowing supercurrent, we
propose a hybrid SQUID-like setup, as indicated in
Fig.~\ref{fig3:exp_setup}, with one arm of a low-$T_c$ material and
the other of the cuprate material of interest. The two arms are in
electrical contact via two Josephson junctions.  Thus, by
construction, the generated flux will mainly originate from the
cuprate arm and we can predict a non-quantized flux of size
\begin{equation}
\frac{\Phi(T)}{\Phi_0} = n2\pi +
\frac{\lambda(T)^2}{\lambda_0^2}
\frac{e \eta(T)}{\sigma_N}
\frac{\delta T}{\Gamma}
+ {\cal {O}}\left[{\sf s}^2\right].
\label{eq:flux}
\end{equation}
Here is $\Phi_0=c_0/2e$ the flux quantum, $c_0$ the velocity of light,
$\delta T=T_1-T_2$ the temperature bias, $\lambda_0=\sqrt{c_0^2/(4\pi
e^2\cNf v_f^2)}$ the zero-temperature penetration depth in the clean
system, and $\lambda(T)=\sqrt{c_0^2m/(4\pi e^2 n_s)}$ the temperature
dependent penetration depth in the dirty system. We assume that the
equilibrium flux is zero and put the integer $n=0$ hereafter. The
superfluid momentum is proportional to the local temperature gradient
$\nabla T$, which leads to that the flux $\Phi$ is proportional to the
temperature bias $\delta T$, since $\Phi$ is related to a contour
integral of $p_s$ around the loop (the evaluation of $\Phi$ is a
standard calculation that can be found in
e.g. Ref.~[\onlinecite{har80}]). Note that if both arms were of the
same material, the thermoelectric response in the two arms would give
counter-flow in opposite directions around the loop and cancel. It is
therefore necessary to have different responses in the two arms,
although it does not matter how that is achieved. Our setup is limited
to low temperatures by the low transition temperature $\sim 10-15 K$
of the $s$-wave superconductor. To probe the high temperature regime
of Eq.~(\ref{eq:flux}), some other type of asymmetry in the
thermoelectric response of the two arms must be
accomplished. Disregarding these complications, we now investigate
Eq.~(\ref{eq:flux}) for all temperatures below $T_c$ of the $d$-wave
arm.

In Fig.~\ref{fig4:Phi} we give an estimate of the induced flux
corresponding to the thermoelectric response in Fig.~\ref{fig2:eta},
assuming a temperature difference of $0.005 T_c \approx 0.5 K$
maintained across the loop. When the scattering phase shift is
intermediate between the Born and unitary limits, fluxes of order of
$0.1\%-1\%$ of the flux quantum is generated at $^4\mbox{He}$
temperatures $\sim 4.2K$ ($\sim 0.05T_c$). At higher temperatures the
temperature difference could be allowed to be larger and we can
predict fluxes on the order of $10\%$ of $\Phi_o$. For cleaner
samples, the induced flux can be even larger, of the order of a few
percent of $\Phi_o$ at $T=0.05\,T_c$, see Fig.~\ref{fig5:Phi_strong}.

For high temperatures, the temperature dependence of the flux is
heavily influenced by the temperature dependence of the penetration
depth, see Fig.~\ref{fig4:Phi} and Fig.~\ref{fig5:Phi_strong}. Thus,
to extract the $T$-dependence of $\eta(T)$ from $\Phi(T)$, the
temperature dependence of $\lambda(T)$ should be divided out. However,
in the important low-temperature regime (to be discussed further in
the following section), $\Phi$ is directly proportional to
$\eta(T)\propto T$, since the penetration depth is limited by impurity
scattering,\cite{hir93b,xu95} $\lambda(T)-\lambda(0)\propto T^2$, and
its $T$-dependence can be neglected.

We note that the minimum temperature at which measurements can be
performed will be set by a combination of two factors: first, the
smallest temperature bias that can be applied ($\delta T\ll T$ is
needed in order to have a uniform thermoelectric response function
$\eta(T)$ in the sample); second, the flux measurement sensitivity,
since the flux scales as $\Phi\sim\eta(T)\delta T\sim T\delta
T$. Thus, at low temperatures, there is a trade off between having a
small $\delta T$ and at the same time have a measurable flux.

Note that the induced superfluid momentum $p_s$ is small. We estimate
\begin{equation}
\frac{v_fp_s}{T_c} \simeq \frac{\xi_0}{L} \frac{\Phi}{\Phi_0},
\end{equation}
where $L$ is the distance between the two reservoirs (i.e. we have set
$\nabla T=\delta T/L$). Thus, we do not need to take feedback effects
from the superflow via the Doppler shifts into account.

\section{Low temperatures}\label{sec:lowT}

\begin{figure*}[t]
\includegraphics[width=16cm]{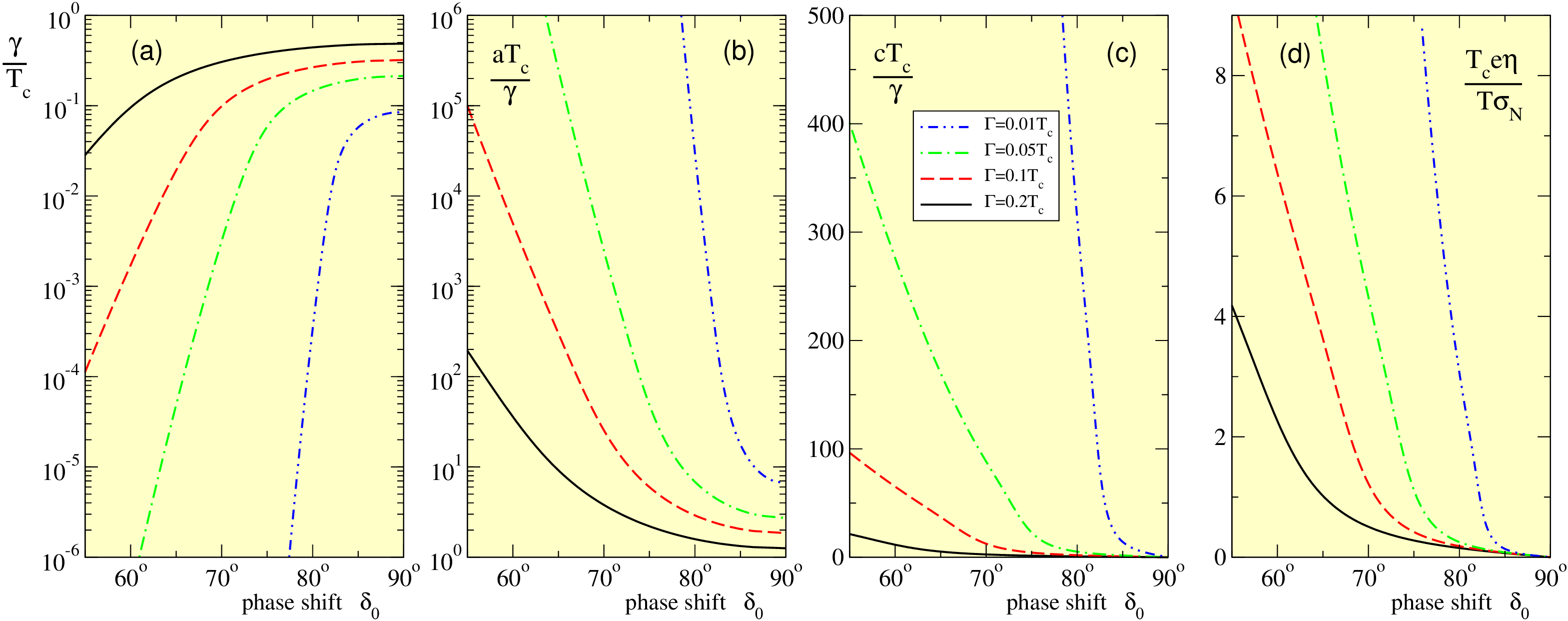}
\caption{(a)-(c) Phase-shift dependence of the impurity band width
$\gamma$ and the parameters $a$ and $c$ normalized by $\gamma/T_c$,
see Eqs.~(\ref{eq:gamma})-(\ref{eq:ac}). (d) The thermoelectric
coefficient $\eta/T$ at low temperatures $T\ll\gamma$. The dashed
(red) curve contains the low-$T$ constants in Fig.~\ref{fig2:eta}. We
have used $\Delta(\vpf)=\Delta_0\cos 2\phi_{\vpf}$
throughout.}\label{fig6:LowE}
\end{figure*}
In the low-$T$ limit, the response functions $\L_{\alpha\beta}$ can be
studied analytically through a systematic Sommerfeld expansion in the
small parameter $T/\gamma$. The magnitude of $\gamma$ depends on the
normal state scattering rate as well as the phase shift, and we have
$\gamma\sim \sqrt{\pi \Delta_0 \Gamma/2}$ in the unitary limit while
$\gamma\sim 4 \Delta_0 \exp[-\pi\Delta_0/2 \Gamma]$ in the Born
limit. For intermediate phase shifts $\gamma(\Gamma,\delta_0)$ can be
found numerically as we discuss below. Only in the limit $T\ll\gamma$
is the Sommerfeld expansion useful. Thus, we consider low energies and
write
\begin{equation}\begin{split}
\tilde \epsilon^R &= a\,\epsilon + i(\gamma + b\,\epsilon^2) 
+ d\,\epsilon^3 + {\cal O}[\epsilon^4],\\
\Im \Sigma^R_0 &= c\,\epsilon + h\,\epsilon^3 + {\cal O}[\epsilon^5],
\end{split}\label{lowTansatz}\end{equation}
with real constants $a$, $b$, $c$, $d$, $h$, and $\gamma$. These
constants are determined self-consistently by the low-energy equations
presented in the appendix. Inserting this ansatz into
Eq.~(\ref{eq:L12}), keeping terms to leading order in $\epsilon$, we
find for the thermoelectric response
\begin{equation}
\frac{\eta(T\ll\gamma)}{T} \simeq 
e\,\frac{\pi^2}{3} \frac{2 \cN_f v_f^2}{\pi \mu \Delta_0}
\left(\frac{c}{\gamma}\right) \left(
1 + \frac{7\pi^2}{15} \frac{a_{12}^2T^2}{\gamma^2}\right),
\label{eq:eta-lowT}
\end{equation}
were $\mu=(1/\Delta_0)|d\Delta(\phi)/d\phi|_{\phi_{node}}$ is the
opening rate of the gap at the nodes (related to the so called gap
velocity, see Ref.~[\onlinecite{durst00}]). The coefficient of the
$T^2$ correction to the $T\rightarrow 0$ asymptotic is given by
\begin{equation}
a_{12}^2 = a^2+2c^2+\frac{3h\gamma^2}{c}-3b\gamma.
\end{equation}
Including phase shifts away from the unitary and the Born limits will
also change the rate at which the charge and heat conductivities
approach their universal values. We find the leading order dependence
of the charge conductivity to be
\begin{equation}
\sigma(T\ll\gamma,\omega\rightarrow 0) \simeq 
e^2\frac{2 \cN_f v_f^2}{\pi \mu \Delta_0}
\left(1+\frac{\pi^2}{3}\frac{a_{11}^2T^2}{\gamma^2}\right),
\label{eq:sigma-lowT}
\end{equation}
and of the heat conductivity to be
\begin{equation}
\frac{\kappa(T\ll\gamma)}{T} \simeq 
\frac{\pi^2}{3}\frac{2 \cN_f v_f^2}{\pi \mu \Delta_0}
\left(1 + \frac{7\pi^2}{15}\frac{a_{22}^2T^2}{\gamma^2}\right),
\label{eq:kappa-lowT}
\end{equation}
where the coefficients of the $T^2$-terms contain direct information
about the impurity induced particle-hole asymmetry through the
parameters $a$ and $c$,
\begin{equation}\begin{split}
a_{11}^2 &= a^2+c^2,\\
a_{22}^2 &= a^2+2c^2.
\end{split}\end{equation}
In the zero temperature limit $T\rightarrow 0$, both the charge and
heat conductivities approach universal values (independent of the
properties of the impurities) while $\eta/T$ has a non-universal
$T\rightarrow 0$ limit set by the ratio $c/\gamma$. All functions,
including $\eta$, have non-universal low-$T$ corrections to their
$T=0$ values and are sensitive to the phase shift $\delta_0$ and the
scattering rate $\Gamma$. We note that the low-$T$ expressions given
in Ref.~[\onlinecite{graf96}] for $\sigma$ and $\kappa/T$ were
implicitly restricted to the unitary and Born limits. Consequently,
the parameter $c$ is absent in their Eqs.~(56)-(57).

\begin{figure*}[t]
\includegraphics[width=16cm]{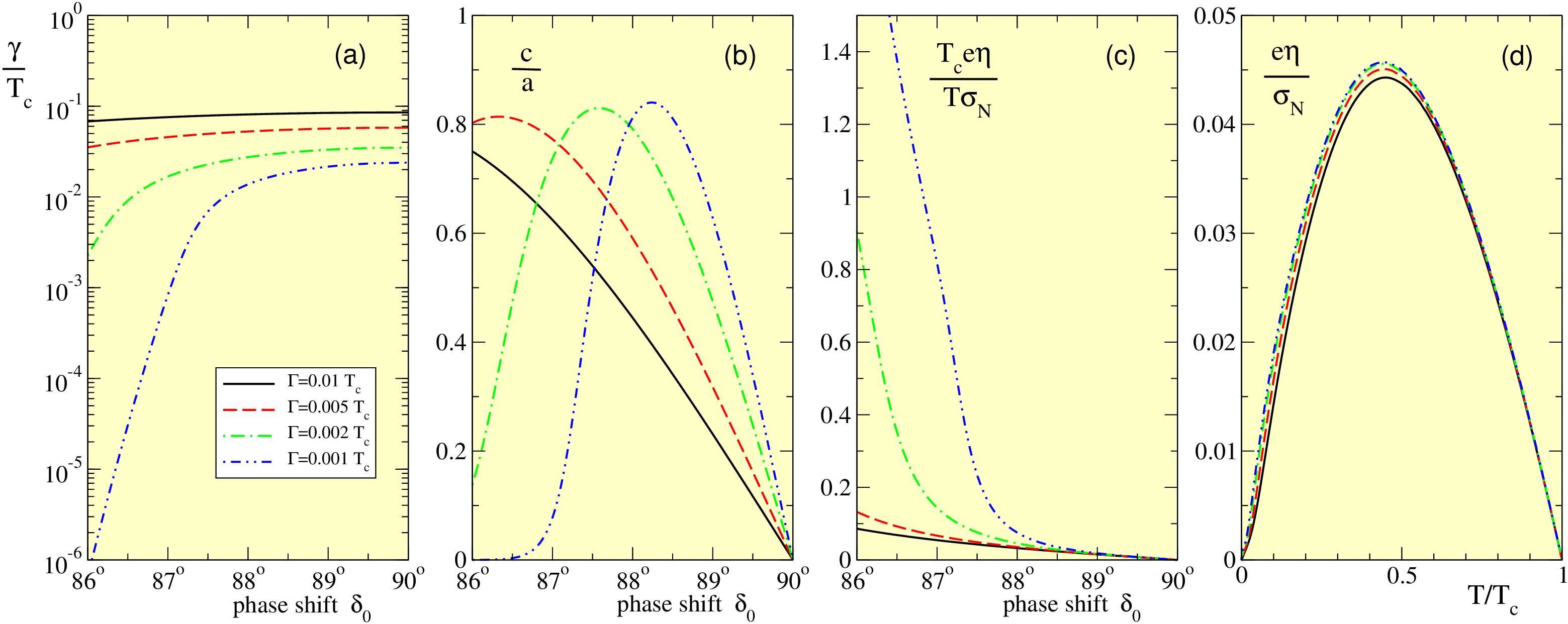}
\caption{(a) The impurity band width $\gamma$ is extremely sensitive
to phase shift variations in clean systems. (b) The ratio of the
parameters $c$ and $a$. (c) The low-$T$ asymptotic of the
thermoelectric coefficient grows large even for small deviations of
the phase shift from the unitary limit. (d) The temperature dependence
of the thermoelectric response function for a phase shift
$\delta_0\approx 86.4^o$ ($\sigma_i=\sin^2\delta_0=0.996$). Note the
data collapse when $\eta$ is scaled by $\sigma_N/e$, because $\eta$
grows as $1/\Gamma$ in clean systems.}
\label{fig7:LowEclean}
\end{figure*}

We can analyze these results further by solving the self-consistency
equations for the impurity self energy to lowest order in $\epsilon$,
and thereby determine the parameters $\gamma$, $a$, $c$, etc. The
results are given in the appendix,
Eqs.~(\ref{eq:gamma})-(\ref{eq:bdh}), and the numerical solution is
presented in Figs.~\ref{fig6:LowE}-\ref{fig7:LowEclean}. The impurity
band width $\gamma$ is exponentially small for phase shifts far from
the unitary limit. As a consequence, the universal limit is reached at
an exponentially small temperature. This is also confirmed in the left
half of Fig. \ref{fig2:eta}, where we present the thermoelectric
response function on a logarithmic scale. Clearly, the crossover to
the low-$T$ power law is severely pushed down in $T$ for small phase
shifts. This suppression becomes extremely fast when the system is
clean. In fact, for ultra-clean samples, with $\Gamma\sim 10^{-3}T_c$
or lower (corresponding to normal state mean free paths $\ell\sim
500\xi_0$ or longer), a deviation of the phase shift away from $\pi/2$
by only a few percent will reduce $\gamma$ by several orders of
magnitude, see Fig.~\ref{fig7:LowEclean}(a). At the same time, the
ratio $c/\gamma$ is increased humongously [Fig.~\ref{fig6:LowE}(c)],
$\eta/T$ grows large [Fig.~\ref{fig7:LowEclean}(c)] and the
thermoelectric coefficient itself is sizable on the scale of
$\sigma_N/e$ [Fig.~\ref{fig7:LowEclean}(d)].

The thermoelectric response scales as $1/\Gamma$, which is also clear
in Fig.~\ref{fig5:Phi_strong}. Thus, if we change the normalization in
Fig.~\ref{fig7:LowEclean}(c)-(d) from the normal state Drude
conductivity $\sigma_N$ (which effectively cancels this scaling), to
the universal limit charge conductance $\sigma(T=0)=\sigma_{00}$ in
Eq.~(\ref{eq:sigma-lowT}), we get the scale $T_c
e\eta/(T\sigma_{00})\sim 100-1000$ in Fig.~\ref{fig7:LowEclean}(c),
and $e\eta/\sigma_{00}\sim 1-10$ in Fig.~\ref{fig7:LowEclean}(d), for
$\Gamma\sim 0.01-0.001T_c$. Thus, the thermoelectric effect grows
large in clean systems with phase shifts close to the unitary limit.

Note the equal importance of the parameters $c$ and $a$ in a region
near the unitary limit, see Figs.~\ref{fig6:LowE}(b)-(c) and
Fig.~\ref{fig7:LowEclean}(b). The ratio approximately behaves as
$c/a\sim (\gamma/\Gamma)\sin\delta_0\cos\delta_0$. Thus, unless
$\gamma\ll\Gamma$ (which holds far from the unitary limit deep inside
the region where $\gamma$ is exponentially suppressed) it is in
general not allowed to neglect the unit term when conductivities are
computed at intermediate phase shifts, although that has been common
in the literature (e.g. in [\onlinecite{hir88,durst00,kim03}]).

Finally, by comparing Figs.~\ref{fig6:LowE} and \ref{fig2:eta} we see
that the true universal limit is in fact reached at very small
temperatures, $T\alt 10^{-2}\gamma$. This is due to the largeness of
the parameters $a/\gamma$ and $c/\gamma$
[c.f. Fig.~\ref{fig6:LowE}(b)-(c)] and higher order parameters $b$,
$d$, and $h$, for clean systems with phase shifts near but not
strictly equal to $\pi/2$. Thus, we need $T\alt(\gamma/a)\ll\gamma$ in
order for the Sommerfeld expansion to work well. Or, strictly
speaking, we have an effective Sommerfeld expansion parameter
$T/(\gamma/a)$, instead of $T/\gamma$.

\section{Summary and conclusions}

In this paper we have focused on phase shifts close to $\pi/2$,
because that appears to be the relevant region experimentally.
Perhaps the most clear signature of this is the impurity induced
resonances seen in STM experiments.\cite{pan00} The resonance energy
is near the Fermi level, but not at the Fermi level, thus signaling a
phase shift close to $\pi/2$. Also, if the low temperature heat
conductivities found in
experiments\cite{tai97,chi00,nak01,and02,pro02,sut03,hill04} are
indeed universal, the phase shift needs to be close to
$\pi/2$. Otherwise, $\gamma$ is exponentially suppressed $\gamma\ll
\Gamma$ and it would be hard to argue that the limit $T\ll\gamma$ was
reached. The limit $\delta_0=\arctan(\pi\cN_fu_0)\rightarrow \pi/2$
can in real life of course only be approached asymptotically. It is
therefore likely that thermoelectric effects are of importance in the
high-$T_c$, as well as other unconventional superconductors (universal
heat conductance was also found in ${\rm Sr}_2{\rm RuO}_4$
[\onlinecite{suz02}]). The effect grows large in clean systems. The
ideal experiment would then be to measure the thermoelectrically
induced flux, together with the thermal conductivity, at low
temperatures for samples with different levels of disorder, e.g
through controlled ${\rm Zn}$ doping. The slope of the thermal
conductivity is related to the gap size $\Delta_0$ and its slope at
the node $\mu$, through $(\pi^2/3)(2\cN_f v_f^2)/(\pi\mu\Delta_0)$,
which can be compared with values of $\Delta_0$ and $\mu$ from other
experimental probes (e.q. photoemission\cite{dam03}). The ratio
$\eta/(e\kappa)=c/\gamma$ contains information about $\Gamma$ and
$\delta_0$ through Eqs.~(\ref{eq:gamma})-(\ref{eq:ac}), with
$\Delta_0$ and $\mu$ as input parameters. We note that the exact angle
dependence of the gap function is not particularly important and it is
enough to know $\Delta_0$ and $\mu$. If the normal state scattering
rate $\Gamma$ is known independently, the scattering phase shift is
uniquely determined by $\eta/(e\kappa)$.

When the energy scale $\gamma$ is exponentially suppressed in clean
systems, the temperature region in which the $T^2$ term in the
expansions in Eq.~(\ref{eq:eta-lowT})-(\ref{eq:kappa-lowT}) is of
importance, is also suppressed. Thus, it is not necessarily ideal to
study super-clean samples. Another issue with super-clean samples is
the validity of the homogeneous scattering model with only $s$-wave
scattering, in which a dilute concentration of point impurities are
assumed to be randomly distributed in the sample. For sufficiently
clean samples this model has to break down and should be replaced by
more realistic models of extended impurities (suggestions can be found
in e.g. Refs.~[\onlinecite{ada02,she03,pog04}]).

In conclusion, we have computed the thermoelectric effect in
unconventional superconductors with impurities scattering in neither
the Born nor the unitary limits. The thermoelectric effect is an
interesting unexplored avenue for the investigation of microscopic
properties of high-$T_c$ as well as other unconventional
superconductors, in particular at low temperatures. Of special
interest is to extract material parameters such as the gap size
$\Delta_0$, the slope of the gap at the gap node
$\mu=(1/\Delta_0)|d\Delta(\phi)/d\phi|_{\phi_{node}}$, the impurity
band width $\gamma$, and the impurity scattering rate $\Gamma$ and
phase shift $\delta_0$. This can be accomplished by measuring the
universal values of the transport coefficients $\sigma$ and $\kappa$
at $T\rightarrow 0$, and possibly their low-$T$ corrections. We add to
this arsenal of tools the thermoelectric coefficient, which at low
temperatures scales linearly with temperature, with a non-universal
slope.

\begin{acknowledgments}
We gratefully acknowledge valuable discussions with V. Chandrasekhar,
P. Delsing, M. Eschrig, Yu. Gal'perin, Z. Jiang, J. A. Sauls,
V. S. Shumeiko, A. Vorontsov, and A. Yurgens. Financial support was
provided by STINT, the Swedish Foundation for International
Cooperation in Research and Higher Education (T.L), the Wenner-Gren
Foundations (T.L.), and the Swedish Research Council (M.F.).
\end{acknowledgments}

\appendix

\section{Impurity self-energy at low energies}

The ansatz in Eq.~(\ref{lowTansatz}) can be used together with the
expression for the self energy in Eq.~(\ref{eq:sR0}) to
self-consistently compute the parameters $a$, $b$, $c$, $d$, $h$, and
$\gamma$. Actually, only $\gamma$ need to be computed
self-consistently; the other parameters immediately follows. To lowest
order, $\epsilon^0$, we obtain
\begin{equation}
\gamma = \frac{\Gamma J_1}{1-\sigma_i+\sigma_i\,J_1^2},
\label{eq:gamma}
\end{equation}
where $J_1$ is the first of a series of Fermi surface integrals
\begin{equation}
J_n=J_n(\gamma) = \int d\vp_f
\frac{\gamma^{n}}{\left[\Delta^2(\vp_f)+\gamma^2\right]^{n/2}},
\end{equation}
where $n$ is an integer (only odd $n$ appears here). In the next
order, $\epsilon^1$, we get
\begin{equation}\begin{split}
a &= \Gamma J_1 D^{-1},\\
c &= 2 \gamma \sqrt{\sigma_i(1-\sigma_i)}(J_1-J_3) D^{-1},
\end{split}\label{eq:ac}\end{equation}
where $D=2\sigma_i\gamma J_1(J_1-J_3) + \Gamma J_3$. The next order,
$\epsilon^2$, is more complicated, but can be expressed in terms of
the following functions
\begin{equation}\begin{split}
k_0 &= J_1,\\
k_1 &= \frac{a}{\gamma}(J_1-J_3),\\
k_2 &= \frac{b}{\gamma}(J_1-J_3) + \frac{3}{2}\frac{a^2}{\gamma^2}(J_3-J_5),\\
k_3 &= \frac{d}{\gamma}(J_1-J_3) +
       \left(\frac{1}{2}\frac{a^3}{\gamma^3}-\frac{3ab}{\gamma^2}\right)(J_3-J_5)\\
    &\quad - \frac{5}{2}\frac{a^3}{\gamma^3}(J_5-J_7),
\end{split}\end{equation}
and
\begin{equation}\begin{split}
m_0 &= 1-\sigma_i+\sigma_i k_0^2,\\
m_1 &= -\frac{2\sigma_i}{m_0}k_0k_1,\\
m_2 &= -\frac{\sigma_i}{m_0}(k_1^2-2k_0k_2),\\
m_3 &= -\frac{2\sigma_i}{m_0}(k_0k_3+k_1k_2).
\end{split}\end{equation}
We get
\begin{widetext}
\begin{equation}\begin{split}
b &= D^{-1} \left\{ 2\sigma_i J_1(J_1-J_3) a(a-1)
  + \frac{a^2}{\gamma} \left[
  \frac{3}{2}(J_3-J_5)(\Gamma-2\sigma_i\gamma J_1)
  + \sigma_i\gamma(J_1-J_3)^2\right]\right\},\\
d &= \gamma D^{-1} \left\{ (\Gamma-2\sigma_i\gamma J_1) \left[
  \left(\frac{1}{2}\frac{a^3}{\gamma^3}-\frac{3ab}{\gamma^2}\right)(J_3-J_5)
  - \frac{5}{2}\frac{a^3}{\gamma^3}(J_5-J_7) \right]
  + m_0m_1b + m_0m_2(1-a)-2\sigma_i\gamma k_1k_2 \right\},\\
h &= \frac{\Gamma\sqrt{1-\sigma_i}}{\sqrt{\sigma_i}m_0}\left[m_1^3+2m_1m_2-m_3\right],
\end{split}\label{eq:bdh}\end{equation}
\end{widetext}

The Fermi surface integrals can in general be computed through
recursion ($n=1$, $3$, $5$, ...),
\begin{equation}
J_{n+2}=J_n - \frac{\gamma}{n}\frac{dJ_n}{d\gamma}.
\end{equation}
To get explicit expressions we need a model of the order
parameter. For the numerics in this paper we use for simplicity
$\Delta(\vpf)=\Delta_0\cos(2\phi_{\vpf})$, in which case $J_1 =
(2/\pi)\sqrt{1-k^2} K\left[k\right]$, where
$k=1/\sqrt{1+(\gamma/\Delta_0)^2}$ and $K\left[k\right]$ is the
complete elliptic integral of the first kind.\cite{GR} In this case
$J_3=(2/\pi)\sqrt{1-k^2} E\left[k\right]$, where $E\left[k\right]$ is
the complete elliptic integral of the second kind, and the other $J_n$
follow by recursion.


\end{document}